\documentclass[a4paper, 11pt, superscriptaddress, prd, showpacs, 
amsmath, amssymb, nofootinbib, showkeys]{revtex4-1}

\usepackage[german,english]{babel}
\usepackage{hyperref}
\usepackage{bm}
\usepackage{bbm}
\usepackage{braket}
\usepackage{slashed}
\usepackage{multirow}
\usepackage{epsfig}
\usepackage{epstopdf}

\newcommand{\js}{J/\psi}
\newcommand{\e}{\eta_c}
\newcommand{\g}{\gamma}

\usepackage{xcolor}

\begin{document}

\title{Line shape and the experimental determination \\ of the 
\texorpdfstring{$\js\to \g\e$}{J/psi->eta gamma} branching fraction}

\author{Jorge Segovia}
\email[]{jsegovia@upo.es}
\affiliation{Departamento de Sistemas F\'isicos, Qu\'imicos y Naturales, Universidad Pablo de Olavide, E-41013 Sevilla, Spain}

\author{Jaume Tarr\'us Castell\`a}
\email{jtarrus@ifae.es}
\affiliation{Grup de F\'\i sica Te\`orica, Departament de F\'\i sica and IFAE-BIST, Universitat Aut\`onoma de Barcelona,\\ 
E-08193 Bellaterra (Barcelona), Catalonia, Spain}

\date{\today}
\begin{abstract}
The width $\js\to \g\e$ is often determined from the photon spectrum line shape of the $\js\to \g\e\to \g X$ decay process. We study this line shape in a nonrelativistic effective field theory framework of quantum chromodynamics including the finite width of the $\e$ and include $\mathcal{O}(v^2)$ corrections. We observe that the photon spectrum line shape is divergent at large energies due to polynomially and logarithmically divergent terms, and, therefore, a subtraction and renormalization scheme is needed to integrate over the photon energies in order to obtain the decay width.  We propose to subtract these divergences at the line-shape level in a manner consistent with the calculation of the width in dimensional regularization and $\overline{\hbox{MS}}$ scheme. We analyze CLEO's data with the proposed subtracted line shape and find that the discrepancy between the theoretical prediction and the experimental result is resolved.
\end{abstract}


\maketitle



\section{Introduction}

Heavy quarkonium systems are nonrelativistic bound states characterized by three well-separated scales
\begin{equation}
m \gg p \sim 1/r \sim mv  \gg E \sim mv^2 \,,
\end{equation}
with $m$ the heavy quark mass, $p$ the relative momentum of the heavy quarks and $E$ the bound state energy. The heavy quark velocity, $v$, is assumed to be $v\ll1$. This is reasonably fulfilled in bottomonium ($v^{2} \sim 0.1$) and at least to a certain extent in charmonium ($v^{2}\sim 0.3$). Integrating out these scales, effective field theories (EFTs) of QCD for quarkonium systems can be constructed. Integrating out $m$ produces nonrelativistic QCD (NRQCD)~\cite{Caswell:1985ui,Bodwin:1994jh}, and integrating out $\sim m v$ potential NRQCD (pNRQCD)~\cite{Pineda:1997bj,Brambilla:1999xf}. The latter is particularly well suited for a model-independent description of quarkonium bound states.

The specific details on the construction of pNRQCD depend on the relative size of the $mv$ scale with respect to $\Lambda_{\rm QCD}$; for $m v\gg \Lambda_{\rm QCD}$ we have the weak-coupling regime and for $mv \sim \Lambda_{\rm QCD}$ the strong-coupling regime. In the weak-coupling version of pNRQCD the potential can be computed in perturbation theory and it allows for a clear theoretical analysis of quarkonium observables. In the charmonium sector, the allowed magnetic dipole (M1) transition $\js\to \g\e$ is a particularly good candidate to be well described by weakly-coupled pNRQCD for two main reasons: It involves the two lowest-lying charmonium states, and, unlike electric dipole (E1) transitions, M1 decays are not particularly sensitive to the long-range tail of the quarkonium wave functions. In Ref.~\cite{Brambilla:2005zw}, allowed and hindered M1 transitions for low-lying quarkonium states were studied in weakly-coupled pNRQCD including relativistic and multipole expansion corrections up to $k_{0}^3 v^2/m^2$ precision, with $k_{0}\simeq m_{\js}-m_{\e}$ the photon energy. In Ref.~\cite{Pineda:2013lta}, the determination of the same transitions was improved by incorporating exactly the static potential into the leading-order Hamiltonian and by resuming large logarithms associated with the heavy quark mass scale.

The $\js\to \g\e\to \g X$ branching fraction, where $X$ are the analyzed decay modes of the $\e$, was first measured in $1986$ by the Crystal Ball Collaboration in the inclusive photon spectrum and the value ${\cal B}_{1S}\equiv{\cal B}(J/\psi\to\g\eta_{c}\to\g X)=(1.27\pm0.36)\%$ was obtained. The transition was not measured again until $2009$ by the CLEO Collaboration~\cite{Mitchell:2008aa}, which analyzed all reported $\e$ decay modes except the $p\bar{p}$ one due to its small rate. The CLEO Collaboration measured ${\cal B}_{1S}=(1.98\pm0.09\pm0.30)\%$. More recently the KEDR Collaboration~\cite{Anashin:2010nr, Anashin:2014wva} measured the transition using the inclusive photon spectrum and reported ${\cal B}_{1S}=(3.40\pm0.33)\%$~\cite{Anashin:2014wva}\footnote{Note that this value is not taken into account into the Particle Data Group (PDG) average~\cite{Zyla:2020zbs}, which considers only the Crystal Ball and CLEO measurements. We also requested the data to include in the analysis carried out in this paper but it was not made available to us.}. As we will show in this paper, these branching ratios, have been incorrectly reported by Refs.~\cite{Mitchell:2008aa,Anashin:2010nr} as corresponding to the $\js\to \g\e$ transition when, in fact, they refer to the $\js\to \g\e\to \g X$ process. The difference between the two processes was partially recognized in Ref.~\cite{Anashin:2014wva}, where the peak of the photon spectrum line shape for the $\js\to \g\e\to \g X$ process was related to $\js\to \g\e$ width. We will expand on this idea and show how to obtain the $\js\to \g\e$ width from the $\js\to \g\e\to \g X$ one.

One of the crucial ingredients in the determination of the branching fraction from experimental measurements is the photon spectrum line shape used in the analysis. The line shape is fitted together with the background to the experimental data, and the number of events above background is used to determine the branching fraction. The CLEO Collaboration in Ref.~\cite{Mitchell:2008aa} observed for the first time a clear asymmetry in the photon energy spectrum line shape due to phase-space- and energy-dependent terms in the $\js \to\g\e$ transition matrix element~\cite{Brambilla:2005zw}. In order to obtain a good fit to the data, the photon spectrum line shape was constructed with a relativistic Breit-Wigner distribution modified by a factor $k^3$, where $k$ is the photon energy. However, adding this factor led to a divergent tail at large photon energies. In order to suppress this behavior, an {\it ad hoc} damping function was included, arguing that it modeled the overlap of the charmonia wave functions. Nevertheless, such damping a factor does not appear in the theoretical studies of Refs.~\cite{Brambilla:2005zw, Pineda:2013lta}, and, thus, it is not well justified. The analysis by the KEDR Collaboration~\cite{Anashin:2010nr, Anashin:2014wva} followed a similar approach incorporating a different, non-theoretically-motivated, damping function. 

A line shape for the photon spectrum of the $\js\to \g\e\to \g X$ decay process obtained from weakly coupled pNRQCD, incorporating a finite width for the $\e$, was presented in Ref.~\cite{Brambilla:2010ey}. This photon spectrum line shape was fitted together with the background to CLEO's data and used to determine the $\e$ mass and decay width, yielding, within uncertainties, equivalent results for the decay width but a slightly higher mass (see Table~\ref{ept}). The line shape in Ref.~\cite{Brambilla:2010ey} also presented a divergent tail at large energies. 

The origin of the divergent tail of the pNRQCD line shape can be traced to contributions that, upon integration over the photon energies, produce either polynomial or logarithmic divergences in the decay width. Therefore, to compute the decay width, the regularization and subtraction of these divergences is necessary, which makes it a scheme-dependent quantity. The cancellation of the divergences is achieved by renormalizing operators that give contributions to $\js\to \g X$ that are not included in the $\js\to \g\e\to \g X$ process. We compute the total width regulating the phase space integral using dimensional regularization (DR), which sets the polynomial divergences to zero, and the logarithmic divergence is subtracted in the $\overline{\hbox{MS}}$ scheme. We propose a modified version of the pNRQCD photon spectrum line shape for the $\js\to \g\e\to \g X$ decay in which the terms that originate the UV divergences in the width are subtracted in a manner consistent with the calculation of the decay width in DR and the $\overline{\hbox{MS}}$ scheme, which we detail in Sec.~\ref{sec2}. Using the proposed line shape, we analyze CLEO's data in Sec.~\ref{sec3} and extract the values for the $\e$ mass, width, and the $\js\to \g\e\to \g X$ branching fractions. From the latter, we obtain the value of the $\js\to \g\e$ width from the relation between the two theoretical expressions of the widths. We discuss our results and give some conclusions in Sec.~\ref{sec4}.


\section{Subtracted Line shape and decay width from EFT}
\label{sec2}

The $\js\to\g\e$ transition amplitude has been computed in Ref.~\cite{Brambilla:2005zw} within the weakly coupled pNRQCD approach and can be written, up to $\mathcal{O}(v^2)$, as 
\begin{align}
\mathcal{A}_{\js\to\g\e} = i\frac{e_Qe}{m}\hat{\bm{e}}\cdot(\bm{k}\times\hat{\bm{\epsilon}}^{*} )\left(1+\kappa-\frac{5}{6m^2}\langle \bm{p}^2\rangle\right)\,,\label{m1v}
\end{align}
where $e_Q$ is the electric charge of the charm quark in units of the electron's charge, and $m$ is its mass. The anomalous magnetic moment $\kappa=\alpha_s C_F/(2\pi)$ is evaluated using $\alpha_s(m)=0.3289$\footnote{The value of $\alpha_s(m)$ is obtained with $4$-loop accuracy and $n_{f}=3$ active flavors. We use for the mass of the charm quark $m = m_{c,RS^{\prime}}(0.7\,{\rm GeV}) = 1648\,{\rm 
MeV}$~\cite{Pineda:2013lta}, which represents the charm mass computed in the renormalon subtracted scheme $RS^{\prime}$ at the renormalization scale $\nu_{f}=0.7\,{\rm GeV}$.}. The $\js$ and $\g$ polarization vectors are $\hat{\bm{e}}$ and $\hat{\bm{\epsilon}}$, respectively. The matrix element $\langle \bm{p}^2 \rangle = \langle 1S | \bm{p}^2 | 1S \rangle = 0.4943\,{\rm GeV}^2$ has been computed in Ref.~\cite{Pineda:2013lta}, where the $1S$ wave function is a solution of the leading-order Hamiltonian that includes exactly the static potential up to $\mathcal{O}(\alpha_s^4)$ precision.

\begin{figure}[!t]
\begin{center}
\epsfig{figure=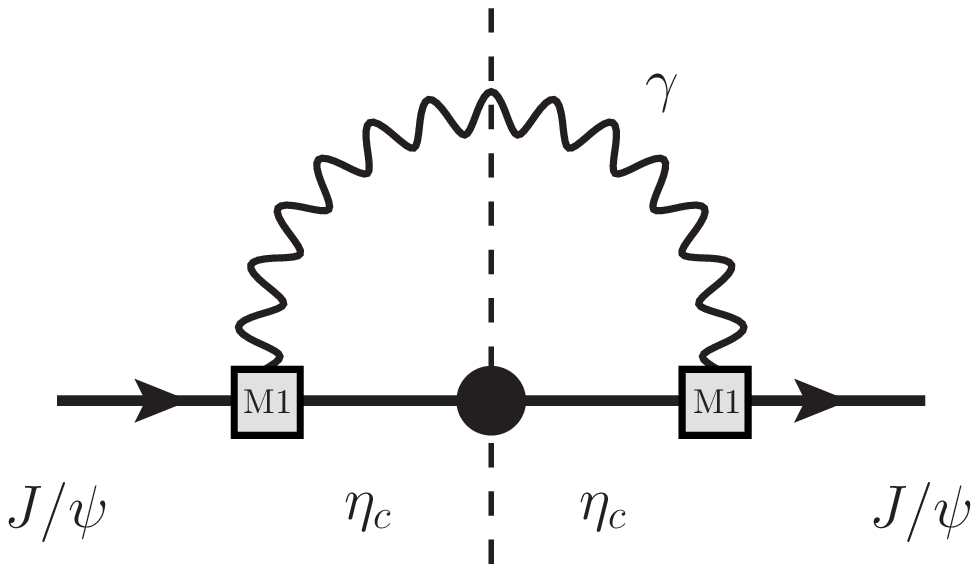,height=0.20\textheight,width=0.60\textwidth}
\caption{\label{lsd} One-loop pNRQCD diagram with M1 vertices relevant for the calculation of the $\js\to\g\e \to\g X$ decay width and line shape. The black dot on the $\e$ propagator represents that a finite width for the $\e$ is incorporated in the propagator. The $\js\to\g\e\to\g X$ decay width is given by the imaginary part of the diagram obtained using the cut represented by the vertical dashed line.}
\end{center}
\end{figure}

The photon spectrum of $\js\to\g\e \to\g X$ can be obtained using the M1 transition vertex in Eq.~\eqref{m1v} by computing the imaginary part of the diagram in Fig.~\ref{lsd} and using the optical theorem. It reads, in DR, as 
\begin{align}
\frac{d\Gamma}{dk}\Big|_{\rm pNRQCD} = 
(4\pi\mu^2)^{\frac{4-d}{2}} \frac{4e^2_Q}{3\pi} \frac{\alpha}{m^2} \left((1+\kappa)^2-\frac{5\langle\bm{p}^2\rangle}{3m^2}\right) k^{d-1} \frac{\Gamma_{\e}/2}{(k-\Delta)^2+\frac{\Gamma^2_{\e}}{4}} \,,
\label{lspnrqcd}
\end{align}
with $\Delta=m_{\js}-m_{\e}$, $d$ the space-time dimension and $\mu$ the renormalization scale. Setting $d=4$ we recover the result in Ref.~\cite{Brambilla:2010ey}. Integrating over all the photon energies, we arrive at
\begin{align}
\Gamma &= 
\frac{e_{Q}^{2}}{3\pi}\frac{\alpha}{m^2}\left((1+\kappa)^2-\frac{5\langle\bm{p}^2\rangle}{3m^2}\right)\Bigg\{\Delta\left(4\Delta^2-3\Gamma_{\e}^2 \right) \left( \pi - \arctan \frac{\Gamma_{\e}}{2\Delta} \right) \nonumber \\
&+\Gamma_{\e} \left(3\Delta^{2}-\frac{\Gamma_{\e}^{2}}{4} \right)\left(\lambda+ 2 -\log\left(\frac{4\Delta^{2}+\Gamma_{\e}^2}{\mu^2}\right) \right) \Bigg\} \,,\label{width}
\end{align}
with $\lambda=2/(4-d)-\g_E+\log 4\pi$. The UV divergence in Eq.~\eqref{width} can be renormalized in the $\overline{\hbox{MS}}$ scheme by absorbing the term proportional to $\lambda$ into suitable operators contributing to the $\js\to\gamma X$ decay. The dependence on the renormalization scale of Eqs.~\eqref{lspnrqcd} and \eqref{width} should be compensated by the scale dependence of the matching coefficients of the same operators. These matching coefficients encode nonperturbative processes and cannot be computed in perturbation theory.

The line shape in Eq.~\eqref{lspnrqcd} is divergent at large photon energies. The terms that produce this behavior can be isolated using partial fractioning or just by looking at the expansion of Eq.~\eqref{lspnrqcd} for large $k$. Two terms are polynomially divergent, and one is logarithmically divergent. In DR, after integrating over the energies, the polynomially divergent terms produce no contribution to the decay width, and the logarithmic divergence is subtracted and canceled by a counterterm. Therefore, the terms responsible for the divergence of the photon spectrum line shape at large energies do not contribute to the decay width. Our proposal is to subtract the divergent terms from the photon spectrum line shape in a manner consistent with the calculation of the decay width in DR and the $\overline{\hbox{MS}}$ scheme. A similar scheme in which the UV divergences are subtracted at the integrand level has been developed in Refs.~\cite{Pittau:2012zd, Pittau:2013ica}. The divergent terms of the photon spectrum line shape read as follows:
\begin{align}
\frac{d\Gamma}{dk}\Big|_{\text{UV div.}} &=(4\pi\mu^2)^{\frac{4-d}{2}}\frac{2e^2_Q}{3\pi}\frac{\alpha \Gamma_{\e}}{m^2}\left((1+\kappa)^2-\frac{5\langle\bm{p}^2\rangle}{3m^2} \right)k^{d-2} \nonumber \\
&\times\left(\frac{1}{k}+\frac{2\Delta}{k^2}+\frac{1}{(k^2+\mu^2)^{3/2}}\left(3\Delta^2-\frac{\Gamma^2_{\e}}{4}\right)\right) \,.\label{lsuvd}
\end{align}
To regularize spurious infrared divergences produced by the separation of the logarithmic divergence, we have introduced a regulator in the last term of Eq.~\eqref{lsuvd}. Setting this regulator to be the same as the renormalization scale cancels the dependence on $\mu$ upon integration of Eq.~\eqref{lsuvd} over the photon energy. We define the subtracted photon spectrum line shape as 
\begin{align}
\frac{d\Gamma}{dk}\Big|_{\text{pNRQCD}_{\text{sub}}} = 
\frac{d\Gamma}{dk}\Big|_{\text{pNRQCD}}-\frac{d\Gamma}{dk}\Big|_{\text{UV div.}} \,.
\label{lspnrqcds}
\end{align}
Integrating the subtracted line shape in Eq.~\eqref{lspnrqcds} over the photon energy we obtain exactly the same width in Eq.~\eqref{width} with the UV-divergent term $\lambda$ subtracted. Since any polynomially divergent term vanishes upon integration in DR, one could, in principle, choose a subtraction different from Eq.~\eqref{lsuvd} by any amount of polynomially divergent terms and it would still produce the same renormalized decay width. Nevertheless, only the subtraction in Eq.~\eqref{lsuvd} leaves the photon spectrum line shape free of any IR and UV divergences.

Finally, one can obtain the expression for the $\Gamma_{\js\to\g \e}$ width, obtained in Refs.~\cite{Brambilla:2005zw, Pineda:2013lta}, by taking the limit $\Gamma_{\e}\to 0$ in Eq.~\eqref{width}:
\begin{align}
\Gamma_{\js\to\g \e}= \frac{4e_{Q}^{2}}{3}\frac{\alpha}{m^2}\left((1+\kappa)^2-\frac{5\langle\bm{p}^2\rangle}{3m^2}\right)\Delta^3\,.\label{dunno}
\end{align}  
We can rewrite Eq.~\eqref{width} (subtracting $\lambda$) in terms $\Gamma_{\js\to\g \e}$ and invert the relations to obtain
\begin{align}
\Gamma_{\js\to\g \e}&=4\pi\Delta^3\Gamma\Bigg\{\Delta\left(4\Delta^2-3\Gamma_{\e}^2 \right) \left( \pi - \arctan \frac{\Gamma_{\e}}{2\Delta} \right) \nonumber \\
& +\Gamma_{\e} \left(3\Delta^{2}-\frac{\Gamma_{\e}^{2}}{4} \right)\left(2 -\log\left(\frac{4\Delta^{2}+\Gamma_{\e}^2}{\mu^2}\right) \right) \Bigg\}^{-1} \,.\label{widthin}
\end{align}  
These expressions allows us to obtain the width $\Gamma_{\js\to\g \e}$ in terms of the quantities $\Gamma$,  $m_{\e}$, and $\Gamma_{\e}$, that can be obtained from the analysis of the experimental data for $\js\to\g\e\to\g X$. Note that the $\mu$ dependence of $\Gamma$ cancels out with the dependence on $\mu$ of the logarithmic term.


\section{Analysis of CLEO's data}\label{sec3}

The CLEO Collaboration~\cite{Mitchell:2008aa} studied the $\psi(2S)\to \gamma\eta_c$ and $\js\to \gamma\eta_c$ magnetic dipole transitions using $2.45\times10^6$ $\psi(2S)$ decays collected with the CLEO-c detector at the Cornell Electron Storage Ring. They extract ${\cal B}_{2S} \equiv {\cal B}(\psi(2S)\to\g\e\to\g X)$ from the $640~{\rm MeV}$ photon transition line visible in the inclusive photon energy spectrum from multihadronic events collected at the $\psi(2S)$ resonance. To measure ${\cal B}_{1S}/{\cal B}_{2S}$, the fraction of event chains 
\begin{align}
&\psi(2S)\to\pi^+\pi^-\js, \quad \,\js\to\g\e, \quad \,\e\to X; \\
&\psi(2S)\to\g\e, \quad \,\e\to X;
\end{align}
is taken. The considered $12$ exclusive $\e$ decay modes are denoted by $X$, which includes all reported $\e$ decay modes except the $p\bar{p}$ one that has a small rate. The branching fraction ${\cal B}_{1S}$ is then obtained as the product of ${\cal B}_{2S}$ and ${\cal B}_{1S}/{\cal B}_{2S}$. For the selection of the exclusive $\js$ decays, the recoil mass of the $\pi^+\pi^-$ is used to select the process $\psi(2S)\to\js\pi^+\pi^-$. The $\mathcal{B}_{1S}$ branching fraction is finally obtained as
\begin{align}
\mathcal{B}^{\rm exp.}_{1S} = \frac{(N_{2S}^{\rm in}/N_{2S}^{\rm ex})N_{1S}^{\rm ex}}{\varepsilon^{\rm in}_{2S}(\varepsilon_{1S}^{\rm ex} /\varepsilon_{2S}^{\rm ex})N_{\psi(2S)} \mathcal{B}_{\pi\pi}} 
\label{expbr}
\end{align}
where $N_{1S(2S)}^{\rm in(ex)}$ and $\varepsilon_{1S(2S)}^{\rm in(ex)}$ are, respectively, the observed number of events and calculated efficiencies of the $\js$ and $\psi(2S)$ in inclusive (in) and exclusive (ex) $\e$ channels. $\mathcal{B}_{\pi\pi}$ is the $\psi(2S)\to\js\pi^+\pi^-$ branching fraction and $N_{\psi(2S)}$ the number of $\psi(2S)$ decays. For our analysis, we take the values obtained by the CLEO Collaboration (see Table~\ref{tab2}) except for the one corresponding to $N_{1S}^{\rm ex}$, which is recalculated using the photon spectrum line shapes in Sec.~\ref{sec2}. 

\begin{table}
\begin{center}
\begin{tabular}{lr}
\hline
\hline
$N_{2S}^{\rm in}/N_{2S}^{\rm ex}$                     & $11.07\pm0.33$ \\
$\varepsilon_{1S}^{\rm ex}/\varepsilon_{2S}^{\rm ex}$ & $0.6515$       \\
$\varepsilon_{2S}^{\rm in}$                           & $56.37~\%$      \\ 
$\mathcal{B}_{\pi\pi}$                                & $(35.04\pm0.07\pm0.77)~\%$ \\
$N_{\psi(2S)}$                                        & $(24.5\pm0.5)\times 
10^6$ \\ 
\hline
\hline
\end{tabular}
\caption{\label{tab2} Yields and efficiencies taken from the CLEO Collaboration analysis in Ref.~\cite{Mitchell:2008aa}. The numerical values of $\mathcal{B}_{\pi\pi}$ and $N_{\psi(2S)}$ are from 
Ref.~\cite{Mendez:2008aa}.}
\end{center}
\end{table}

\begin{figure}[!t]
\begin{center}
\epsfig{figure=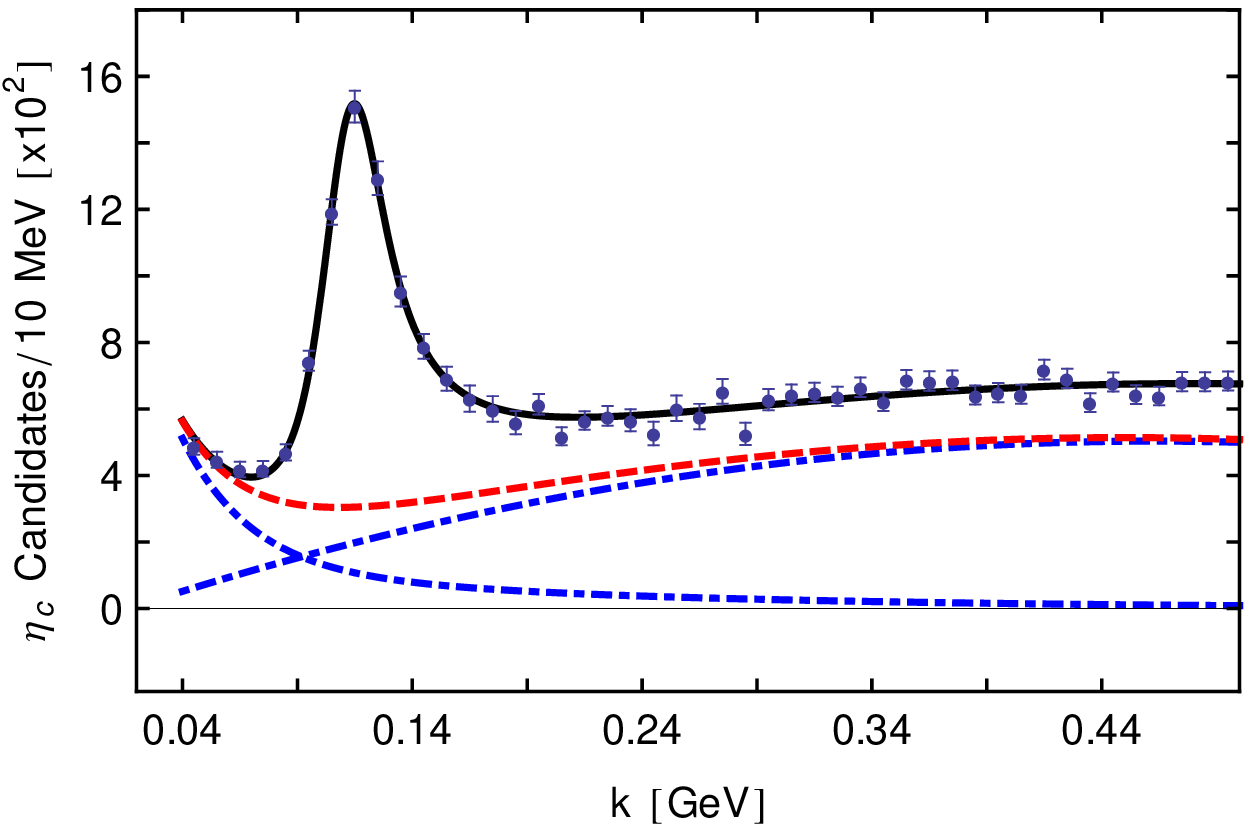,height=0.30\textheight,width=0.45\textwidth}
\put(-35,160){\small (a)}
\hspace*{1.00cm}
\epsfig{figure=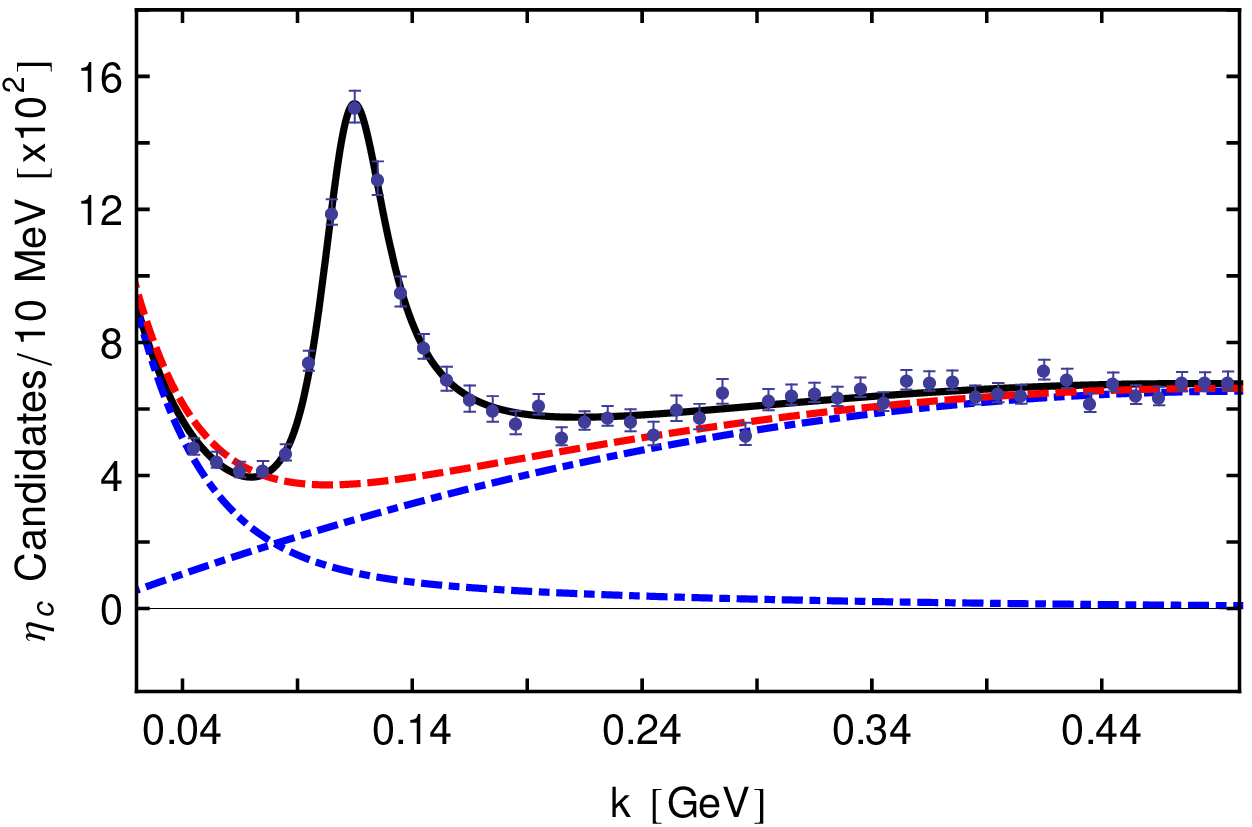,height=0.30\textheight,width=0.45\textwidth}
\put(-35,160){\small (b)}
\caption{\label{fig:LineShape} The (black) solid line is the fit to the photon spectrum in exclusive $J/\psi\to \gamma\eta_{c}\to \gamma X$ decays using the unsubtracted pNRQCD line shape of Eq.~\eqref{lspnrqcd} [panel (a)] and using the subtracted pNRQCD line shape of Eq.~\eqref{lspnrqcds} [panel (b)]. Experimental data points are taken from Ref.~\cite{Mitchell:2008aa}. Total background is given by the (red) dashed line. The (blue) dot-dashed curves indicate the two major background components described in the text.}
\end{center}
\end{figure}

The photon spectrum for the $\js\to\g\e\to\g X$ process measured by the CLEO Collaboration is shown in Fig.~\ref{fig:LineShape}. We have fitted the unsubtracted [Fig.~\ref{fig:LineShape}(a) and Eq.~\eqref{lspnrqcd}] and subtracted [Fig.~\ref{fig:LineShape}(b) and Eq.~\eqref{lspnrqcds}] pNRQCD line shapes with a free normalization together with the background shape. The total signal is convoluted with a resolution function with a width of $4.8~{\rm MeV}$. In both cases, we set $d=4$, take $m_{\js}=3096.9~{\rm MeV}$~\cite{Zyla:2020zbs}, and use $\mu=0.4~{\rm GeV}$ in the subtracted line shape. The remaining parameters, such as charm quark mass, have been specified in Sec.~\ref{sec2}.

The background is the same used by the CLEO Collaboration and it is composed by: (i) a Monte Carlo modeled background for spurious $\js\to X$ with shape
\begin{align}
bkg_1(k)=A_1\left(e^{-5.720k}+10.441e^{-33.567k}\right)\,,
\end{align}
where the normalization $A_1$ is fitted; and (ii) a freely fit background for $\js\to\pi^0 X$ and nonsignal $\js\to \gamma X$ with shape
\begin{align}
bkg_2(k)=B_0+B_1k+B_2k^2\,.
\end{align}

\begin{table}
\begin{center}
\begin{tabular}{c|ccccc}
\hline
\hline
& $m_{\e}$ (MeV) & & $\Gamma_{\e}$ (MeV) & & $\mathcal{B}^{\rm exp.}_{1S}$ 
(\%) \\ 
\hline
pNRQCD & $2985.8\pm0.6$ & & $29.7\pm1.7$ & & $(3.86\pm0.33)$ \\
pNRQCD$_{\text{sub}}$ & $2985.8\pm0.6$ & & $29.7\pm1.7$ & & $(2.17\pm0.18)$ \\
CLEO & $2982.2\pm0.6$ & & $31.5\pm1.5$ & & $(1.98\pm0.09\pm0.30)$ \\
KEDR & $2983.5\pm1.4^{+1.6}_{-3.6}$ & & $27.2\pm3.1^{+5.4}_{-2.6}$ & & 
$(3.40\pm0.33)$ \\
\hline
\hline       
\end{tabular}
\caption{\label{ept} The $\eta_{c}$ resonance parameters: mass and total decay width (in MeV), and the ${\cal B}^{\rm exp.}_{1S}={\cal B}(J/\psi\to \g\eta_{c}\to\g X)$ branching fraction (in percent) computed from the line-shape fit using Eq.~\eqref{expbr}. Using CLEO's data, we quote our results using the unsubtracted and subtracted line shapes, and compare them with those reported by CLEO~\cite{Mitchell:2008aa} and KEDR~\cite{Anashin:2014wva} Collaborations. For comparison purposes, the current PDG~\cite{Zyla:2020zbs} averages for the $\e$ mass and width are $m_{\e}=2983.9\pm0.5$~MeV and $\Gamma_{\e}=32.0\pm0.7$~MeV, respectively.}
\end{center}
\end{table}

In Table~\ref{ept}, we present our results obtained from the analysis of CLEO's data for the $m_{\e}$, $\Gamma_{\e}$, and the branching fraction $\mathcal{B}^{\rm exp.}_{1S}$. The latter is obtained from Eq.~\eqref{expbr} by inputting the value of $N_{1S}^{\rm ex}$ which is obtained as the sum of the number of events above background for the whole range of experimental data. We display our results using the unsubtracted and subtracted line shapes, and compare them with those reported by CLEO~\cite{Mitchell:2008aa} and KEDR~\cite{Anashin:2014wva} Collaborations as well as the PDG averages~\cite{Zyla:2020zbs}. Both unsubtracted and subtracted line shapes yield the same results for the $\e$ mass and decay width. We obtain an $\e$ decay width that is compatible, within uncertainties, with the one reported by CLEO and the PDG average. The difference with respect to the one obtained by KEDR has its origin in the experimental data and not in the line shape used. We obtain an $\e$ mass which is slightly higher than all the experimental measurements, albeit barely within uncertainties.

It can be seen in Fig.~\ref{fig:LineShape} that a substantial tail of signal persists at large photon energies when fitting CLEO's data with the unsubtracted pNRQCD line shape, while using the subtracted line shape removes that tail. As a result, the branching fraction $\mathcal{B}^{\rm exp.}_{1S}$ obtained using the subtracted line shape is $44\%$ smaller than the one obtained with the unsubtracted line shape. The value of the branching fraction $\mathcal{B}^{\rm exp.}_{1S}$ obtained using the subtracted line shape is close to the CLEO's determination, which indicates that the subtraction of the line shape has a similar effect as adding an {\it ad hoc} damping function, however, the latter is not well theoretically motivated. The KEDR result is $\sim55\%$ larger and the difference should be attributed to the experimental data and not to the photon line shape used. Notice that KEDR result is not taken into account in the PDG average.

We can now use the values for the $\e$ mass and decay width in Table~\ref{ept} obtained from our fits to compute the $\js\to\g\e\to\g X$ decay width using the theoretical expression in Eq.~\eqref{width}
\begin{align}
\Gamma=(2.03\pm0.03_{\rm stat.}\pm 0.44_{\rm theo.})\,{\rm keV}\,,\label{eq:GammaTheodr}
\end{align}
where the ${\cal O}(v^2)$ corrections make up about $16\%$ of the total value. The first error corresponds to the statistical uncertainty from the fit, and the second is the theoretical one~\cite{Pineda:2013lta}. The value in Eq.~\eqref{eq:GammaTheodr} corresponds to the branching fraction
\begin{align}
\mathcal{B}^{\rm theo.}_{1S}&\equiv \frac{\Gamma}{\Gamma_{\js}}=(2.19\pm 0.07_{\rm stat.} \pm 0.48_{\rm theo.})\,\% \,,\label{eq:Btheo1}
\end{align}
with $\Gamma_{\js}=92.9\pm2.8~{\rm keV}$~\cite{Zyla:2020zbs}. Now, we can compare this result of the branching ratio with the experimental one in Table~\ref{ept} for the subtracted line shape, and we observe a very good agreement, well within the uncertainties.

A different signal to background ratio is obtained if the renormalization scale $\mu$ is changed. This is consistent with the fact that the renormalization of the logarithmic divergence involves an operator that contributes to a background process. Since the transition amplitude in Eq.~\eqref{m1v} is obtained from pNRQCD, the value of the renormalization scale should be kept of the order of the typical momentum scale of this EFT in order not to spoil the EFT expansion, which is the motivation behind the value $\mu=0.4~{\rm GeV}$ we have used. We have analyzed the sensitivity of $\mathcal{B}^{\rm exp.}_{1S}$ with respect to $\mu$ by modifying it by $10\%$, and found that the branching fraction changes by $\sim3.5\%$. Furthermore, the dependence in $\mu$ is matched by the one in the theoretical expression in Eq.~\eqref{eq:Btheo1} through the $\mu$ dependence of Eq.~\eqref{width}. 

The branching ratio obtained with the unsubtracted line shape is highly dependent on the choice of energy cutoff of the experimental data, while the one obtained with the subtracted line shape produces very similar results for any cutoff above $0.3$~GeV. 

It is interesting to consider the contribution of the subtraction in Eq.~\eqref{lsuvd} integrated for $d=4$ up to a cutoff $\Lambda=495~{\rm MeV}$, which corresponds to the highest photon energy measured by the CLEO Collaboration. The contribution to $\Gamma$ is $1.54~{\rm keV}$, which if added to the determination of the branching fraction gives $\mathcal{B}^{\rm theo.}_{1S}=3.85\%$, coinciding with $\mathcal{B}^{\rm exp.}_{1S}$ determined using the unsubtracted photon spectrum line shape. This indicates that the analysis of the experimental data done using the unsubtracted line shape should be compared with a computation of the width performed with a cutoff regulator.

Finally, the decay width of the $\js\to\g\e$ process can be obtained from the experimental branching ratio of $\js\to\g\e\to\g X$, the $\e$ mass and decay width from Table~\ref{ept} by using Eq.~\eqref{widthin} with the value $\Gamma=\mathcal{B}^{\rm exp.}_{1S}\Gamma_{\js}$. We obtain
\begin{align}
\Gamma^{\rm exp.}_{\js\to\g\e}=(1.82\pm 0.16)~{\rm keV} \,.
\label{eq:Btheo2dr}
\end{align}
We can compare this result with the theoretical expression in Eq.~\eqref{dunno} using the computation of the matrix elements from Ref.~\cite{Pineda:2013lta} and the value of $m_{\e}$ from our fits in Table~\ref{ept}. We find\footnote{The difference with the value in Eq.~\eqref{eq:Btheo} from the one given in Ref.~\cite{Pineda:2013lta} is entirely due to the different $m_{\e}$ used.}
\begin{align}
\Gamma^{\rm theo.}_{\js\to\g\e}=(1.84\pm 0.03_{\rm stat.} \pm 0.40_{\rm theo.})~{\rm keV} \,,\label{eq:Btheo}
\end{align}
which is compatible with the result in Eq.~\eqref{eq:Btheo2dr}. The values obtained in the most recent lattice QCD computations are $\Gamma_{\js\to\g\e}=(2.64\pm11\pm3)~{\rm keV}$~\cite{Becirevic:2012dc} and $\Gamma_{\js\to\g\e}=(2.49\pm19)~{\rm keV}$~\cite{Donald:2012ga}. The first is incompatible with both our experimental and theoretical results, while the second is compatible with only the theoretical determination.


\section{Conclusions}
\label{sec4}

The experimental data on the photon energy spectrum for $\js\to\g\e\to\g X$ have been used in the literature to determine the width of $\js\to\g \e$ with results claimed to be incompatible with the theoretical determinations of Refs.~\cite{Brambilla:2005zw,Pineda:2013lta}. To perform these analyses, an expression of the photon spectrum line shape is needed. This is used to fit the data and evaluate the number of events above the background which can be used to obtain the branching ratio for $\js\to\g\e\to\g X$, which incorrectly has been reported as the branching ratio for $\js\to\g \e$ in some instances.

We have calculated the photon spectrum line shape for the $\js\to\g\e\to\g X$ process in weakly coupled pNRQCD~\cite{Brambilla:2010ey}. We have incorporated to the leading-order expression for the $\js\to\g\e$ transition amplitude the relativistic and multipole expansion corrections up to ${\cal O}(v^2)$~\cite{Brambilla:2005zw}. These corrections have been computed  incorporating the full static potential into the leading-order Hamiltonian as in Ref.~\cite{Pineda:2013lta}. 

We have shown that the large energy tail of the line shape is due to polynomially and logarithmically divergent terms. Because of these divergent terms, the computation of the total width for the $\js\to\g\e\to\g X$ process requires the use of a regularization and subtraction scheme. The counterterms needed for renormalization are provided by other operators contributing to the $\js\to\g X$ process. The scheme dependence of the $\js\to\g\e\to\g X$ width can be understood as resulting from the separation of a subprocess from the total, physically observable, $\js\to\g X$. Integrating the line shape over the photon energy using DR, an analytical expression for the width of the $\js\to\g\e\to\g X$ process has been obtained Eq.~\eqref{width}. Upon integration in DR, the polynomially divergent terms give no contribution, and the logarithmically divergent term produces an UV divergence that is subtracted in $\overline{\hbox{MS}}$ scheme. We have proposed a subtracted photon spectrum line shape, Eq.~\eqref{lspnrqcds}, in which the UV-divergent terms are subtracted in a manner consistent with the calculation of the decay width in DR and the $\overline{\hbox{MS}}$ scheme. 

We have analyzed CLEO's data for $\js\to\g\e\to\g X$ process using the unsubtracted and subtracted line shapes. The signal over background ratio depends on the tail of the line shape at large photon energies. Using the unsubtracted line shape, the large energy tail leads to a determination of ${\cal B}^{\rm exp.}_{1S}$, from Eq.~\eqref{expbr}, which is incompatible with the theoretical determination from Eqs.~\eqref{width} and \eqref{eq:Btheo1}. Using the subtracted line shape in the analysis of the experimental data, the experimental and theoretical determinations of the branching fraction are in good agreement: 
\begin{align}
{\cal B}^{\rm exp.}_{1S} &= (2.17\pm0.18)\% \,, \\
{\cal B}^{\rm theo.}_{1S} &= (2.19\pm 0.07_{\rm stat.} \pm 0.48_{\rm theo.})\% \,.
\end{align}
Both these values depend on the choice of renormalization scale $\mu$, the experimental one through the $\mu$ dependence of the line shape [see Eqs.~\eqref{lspnrqcds} and \eqref{lsuvd}], and the theoretical one through the explicitly $\mu$ dependence of $\Gamma$ in Eq.~\eqref{width}. The $\js\to\g \e$ width can be obtained from Eq.~\eqref{widthin}, where the $\mu$ dependence cancels out, by inputting $\Gamma=\mathcal{B}^{\rm exp.}_{1S}\Gamma_{\js}$. We obtain the following value:
\begin{align}
\Gamma^{\rm exp.}_{\js\to\g\e}=(1.82\pm 0.16)~{\rm keV} \,,
\end{align}
which is compatible with the theoretical result in Eq.~\eqref{eq:Btheo} from Refs.~\cite{Brambilla:2005zw,Pineda:2013lta}.


\bigskip
{\bf Acknowledgements}
\bigskip

We thank Nora Brambilla and Antonio Vairo for making us aware about the discrepancy between the experimental and theoretical determinations of the $\js\to\g\e$ width. J.T.C acknowledges the financial support from the European Union's Horizon 2020 research and innovation program under the Marie Sk\l{}odowska--Curie Grant Agreement No. 665919. He has also been supported in part by the Spanish Grants No. FPA2017-86989-P and SEV-2016-0588 from the Ministerio de Ciencia, Innovaci\'on y Universidades, and the Grants No. 2017-SGR-1069 from the Generalitat de Catalunya. J.S. acknowledges the financial support from Ministerio Espa\~nol de Ciencia e Innovaci\'on under Grant No. PID2019-107844GB-C22; and Junta de Andaluc\'ia,  Contract No. P18-FR-5057 and Operativo FEDER Andaluc\'ia 2014-2020 UHU-1264517.



\bibliography{ls}

\end{document}